\documentclass[fleqn,10pt]{wlscirep}
\title{Straightening of light in a one dimensional dilute photonic crystal}
\author[1,2*]{Zhyrair Gevorkian}
\author[3]{Vladimir Gasparian}
\author[4]{Emilio Cuevas}
\affil[1]{Yerevan Physics Institute, Alikhanian Brothers St. 2, 0036 Yerevan, Armenia}
\affil[2]{Institute of Radiophysics and Electronics, Ashtarak-2, 0203, Armenia}
\affil[3]{California State University, Bakersfield, California 93311-1022, USA}
\affil[4]{Departamento de F\'{i}sica, Universidad de Murcia, E-30071 Murcia, Spain}
\affil[*]{gevork@yerphi.am}

\begin{abstract}
Light transport in a dilute photonic crystal is considered. The analytical expression for the transmission coefficient is derived. Straightening of light under certain conditions in a one-dimensional photonic crystal is predicted. Such behavior is caused by the formation of a localized state in transversal motion. The main contribution to the central diffracted wave transmission coefficient is due to states, that either close to the conductance band's bottom or deeply localized in the forbidden gap.
Both these states suppress mobility in the transverse direction and force light to be straightened. Straightening of light in the optical region along with small reflection make these systems very promising for use in solar cells.
\end{abstract}
\begin{document}

\flushbottom
\maketitle

\thispagestyle{empty}


\section{Introduction}
Recent developments in material science have made possible the fabrication of photonic crystals that allow the observation of many peculiar effects \cite{joann08}, including perfect reflector for all polarizations over a wide selectable spectrum \cite{john,oscar},optical Hall effect \cite{OMN2004}, unidirectional scattering \cite{haldane08} with broken time-reversal symmetry and the propagation of optical beams without spatial spreading. The latter, known as supercollimation, has attracted a lot of attention in various fields of physics. Supercollimation effect has been observed in mesoscopic ($\sim 100 \mu m$) \cite{Kos99, Kos00, wu03, prather04, pustai04, shi04} and macroscopic centimeter-scale photonic crystals \cite{rakich06, miller06}. The latter results indicate that supercollimation effect is very robust and insensitive to possible irregularities or short-scale
disorder in the photonic crystal structure. A standard mechanism that can account for the supercollimation effect of light in two-dimensional photonic crystals is the flat character of dispersion curve $\omega(\vec q)$
in the transverse direction due to the interference of the different transverse
components of the wavevectors (see, for example, Ref. \cite{prather09}).

However, as we will show below, in diluted photonic crystals (DPC) exists another physical mechanism, based on photon's transversal motion restriction, that can lead similar to supercollimation effect \cite{prather09}. The basic idea of this model can be easily understood by considering the following geometry shown in Fig.1. First, suppose that a photon falls on the plates normally on $0x$ direction. In case if the photon's wavenumber $k_x$ lies in the photonic band gap region, the transmission coefficient must be suppressed (see, for example, Ref. \cite{aly12}). Next, imagine that the same photon falls down onto the system
obliquely, as shown in Fig.1 and assume that the transversal to plates component $k_x$ still remains in the photonic band gap. Clearly, for small and moderate scattered angles transversal motion is suppressed due to the appearance of localized or low energy states. Hence, as a direct consequence of the restriction in the transverse direction, photon's propagation parallel to plates will be enhanced.

Note that in most papers on 1d photonic crystal photon main motion is normal to the plates, see, for example, \cite{zhao09} and for recent reviews \cite{review}. In our manuscript we are considering a different geometry (see Fig.1) where photon mainly propagates parallel to the plates. The novelty of our consideration is in different geometry and in corresponding theoretical approach. Our approach allows to obtain closed analytical expression for transmission coefficient.

In this paper we aim to present a complete and quantitative theoretical description of the supercollimation \cite{prather09} and straightening of light \cite{miller06} effects within simplified DPC model, taking into account the above-mentioned restriction in the transverse direction. We will see, that the simplified DPC models, developed in Refs. \cite{GGC16,gagc17}, can help to provide new insights into the properties of the mentioned effects.

The main simplification is related to the ignoring of backscattering in a DPC. Within this approach, we have investigated the transport of light through a one-dimensional metallic photonic crystal with transverse to incident direction inhomogeneity. Independence of transmission coefficient on the incident light wavelength was found\cite{GGC16}. Beside that, we have predicted and experimentally observed \cite{gagc17} a capsize, a drastic change of polarization to the perpendicular direction in DPC. The present work takes one step further in the study of the supercollimation effect. We will remove the limitations on the
normal incident light, discussed in Refs. \cite{GGC16,gagc17} and consider general oblique incidence case. The latter, as we will see below, can lead to an interesting phenomenon of straightening of light into the normal direction on exit of the photonic crystal. In this sense, the DPC can be used in solar cells to increase their efficiency. The point is that at oblique incidence a large amount of light energy is lost due to reflection. Preliminary straightening would reduce this loss. For other applications of 1d photonic crystals see also \cite{zhao2013,peig2015}.

Note, that the main difference of DPC model from other photonic crystal systems where supercollimation effects where observed \cite{Kos99, Kos00, wu03, prather04, pustai04, shi04} is that in the former reflection is negligible due to the fact that the fraction of one component is very small.

\section{Initial Relations}
Consider a system with inhomogeneous dielectric permittivity $\varepsilon(x,y)$ (see Fig.1) and suppose that a plane wave enters the system from the $z < 0$ space at an arbitrary incidence angle $\theta$.
\begin{figure}
 \begin{center}
\includegraphics[width=16.0cm]{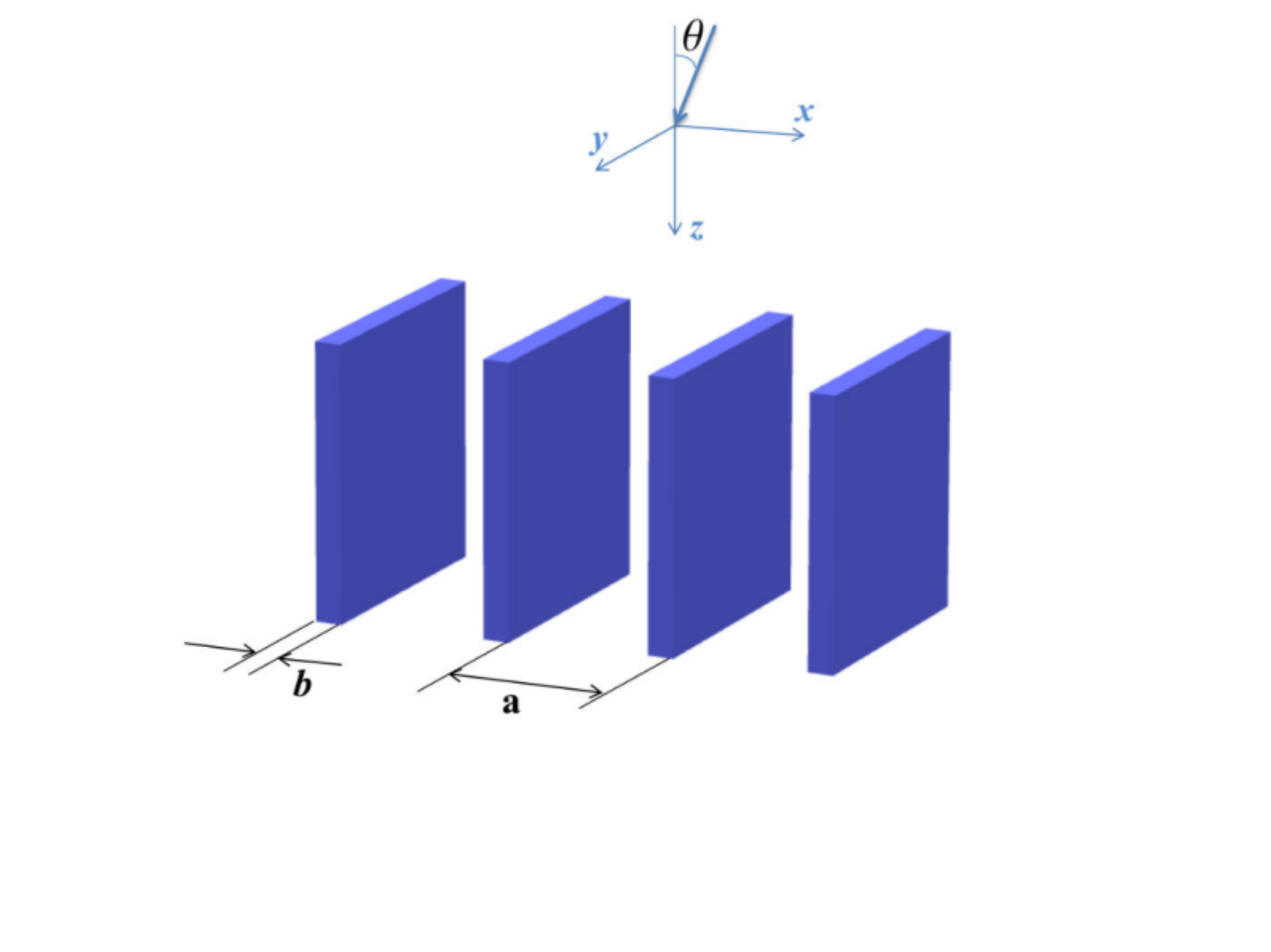}
\caption{Geometry of the problem.}
\label{fig.1}
\end{center}
\end{figure}
Following procedures described in previous work Ref.\cite{GGC16} and \cite{FF79}-\cite{Lag89}, scalar Helmholtz equation is reduced to the following  time-dependent Schr{\"o}dinger equation for a particle with mass $k_0$
\begin{equation}
i\frac{d\phi}{dz}=\hat{H}(x,y)\phi,
\label{sreqtime}
\end{equation}
where
\begin{equation}
\hat{H}(x,y)=-\frac{1}{2k_0}\nabla_t^2+\frac{k_0}{2}\left(1-\varepsilon(x,y)\right).
\label{tem}
\end{equation}
Note, that the parabolic approximation, that is $|d^2\phi/dz^2|<<2k_0|d\phi/dz|$ justified, if the characteristic scale $L_z$ of $\phi(x,y,z)$
along $z$ is much longer than a wavelength, that is
$k_0L_z>>1$ (see below). The solution of Eq.(\ref{sreqtime}) can be represented through the eigenfunctions of Hamiltonian Eq.(\ref{tem}). Therefore for the solution of Helmholtz equation one has
\begin{equation}
\Phi(x,y,z)=e^{ik_0z}\sum_nc_ne^{-iE_nz}\phi_n(x,y).
\label{maxsol}
\end{equation}
where
\begin{equation}
\hat{H}\phi_n(x,y)=E_n\phi_n(x,y).
\label{eigfun}
\end{equation}
It follows from Eq.(\ref{maxsol}) that the local transmission amplitude of a central diffracted wave can be defined as in Ref.\cite{GGC16}
\begin{equation}
t(x,y)=\sum_{E_n<k_0}c_ne^{-iE_nL}\phi_n(x,y),
\label{transamp}
\end{equation}
where $L$ is the system size in the $z$ direction.
\newline
The central diffracted wave transmission coefficient that is measured in the experiment can be estimated by the following expression
\begin{equation}
T=\frac{1}{S}\int dx dy\bigg| t(x,y)\bigg|^2,
\label{trcoeff}
\end{equation}
where $S$ is the area of the system. Substituting Eq.(\ref{transamp}) into  Eq.(\ref{trcoeff}), one has
\begin{equation}
T=\frac{1}{S}\sum_{E_n<k_0}|c_n|^2.
\label{trcoeff2}
\end{equation}
In order to find the coefficients $c_n$ let us consider the Eq.(\ref{maxsol}) for $z=0$
\begin{equation}
\Phi(x,y,z=0)=\sum_nc_n\phi_n(x,y).
\label{zero}
\end{equation}
To proceed, we assume that the wave intruding the system  has an amplitude $1$ (the region $z<0$).
From the continuity at $z=0$, one has $\Phi(x,y,z=0)=e^{k_0sin\theta x}$. Here we ignore the reflected waves from the dilute system. Within this approach, multiplying both sides of Eq.(\ref{zero}) by $\phi_n^*(x,y)$ and integrating over the surface, one has
\begin{equation}
c_n=\int dxdy\phi_n^*(x,y)e^{ik_0sin\theta x}.
\label{expcoef}
\end{equation}
Substituting Eq.(\ref{expcoef}) into Eq.(\ref{trcoeff2}), we arrive at the final result for the transmission coefficient
\begin{equation}
T=\frac{1}{S}\sum_{E_n<k_0}\int d\vec\rho d\vec\rho^{\prime}\phi_n^*(\vec\rho)\phi_n(\vec\rho{\prime})e^{ik_0sin\theta(x-x^{\prime})}, \label{fincoef}
\end{equation}
where $\vec\rho\equiv (x,y)$ is a two dimensional vector on the $xy$ plane.
The equation above is the generalization of the previous result, obtained in Ref. \cite{GGC16}, in case of oblique incidence of light at an arbitrary angle $\theta$. In the succeeding subsections we examine the limitations of the found equation for different models.
\section{Transmission coefficient}
Note that when $k_0<E_b$ ($E_b$ is the bottom value of first energy band), the transmission coefficent (Eq.(\ref{fincoef})) is equal to zero. When $k_0$ is inside the allowed band, transmission coefficient can be represented by quasi-momentum $\vec q$ in the following way
\begin{equation}
 T=\frac{1}{S}\sum_n\int_{E_n(\vec q)<k_0}\frac{d\vec q}{2\pi}\int d\vec\rho d\vec\rho^{\prime}\phi^*_{n\vec q}(\vec\rho)\phi_{n\vec q}(\vec\rho^{\prime})e^{ik_0sin\theta(x-x^{\prime})},
\label{invar}
\end{equation}
where $\vec q$ is integrated over the first Brilloin zone. For sake of simplicity and demonstration of the results, we will carry out further consideration in one-dimensional case.
\subsection{Kronig-Penney model}
Assuming that plates are positioned periodically along the $x$ axis (see Fig.1), we can represent the cross section of the potential as a multitude of square potential wells. Metal layers of PC imitate the potential wells with depth $V_{d}=k_0(\varepsilon-1)/2$ and width $b$. The vacuum layer is charachterized by potential energy $V=0$ and $\varepsilon=1$. So the problem described by Eq.(\ref{tem}) is brought to the Kronig-Penney model \cite{Krpen04}.
The transmission coefficient $T$ for one-dimensional configuration can be written as
\begin{equation}
T=\frac{1}{L_x}\sum_n\int_{E_n(q)<k_0}\frac{dq}{2\pi}\int dx dx^{\prime}\phi_{nq}^*(x)\phi_{nq}(x^{\prime})e^{ik_0sin\theta(x-x^{\prime})}.
\label{onedim}
\end{equation}
where $L_x$ is the system size in the $x$ direction.
Here sum is running over the bands with energy $E_n(\vec q)<k_0$. When $k_0$ lies in the energy gap the integration over $\vec q$ covers whole Brilloin zone $-\pi/a, \pi/a$. Using Bloch theorem $\phi_{nq}(x)=e^{iqx}u_{nq}(x)$, where $u_{nq}(x)$ is a periodical function, one obtains
\begin{eqnarray}
T=\frac{1}{L_x}\sum_n\int_{E_n(q)<k_0}\frac{dq}{2\pi}\sum_{lm}\int_{(l-1)a}^{la}dxe^{-i(q-k_0sin\theta)x}u_{nq}^*(x)\times\nonumber\\
\times \int_{(m-1)a}^{ma}dxe^{i(q-k_0sin\theta)x}u_{nq}(x)
\label{parts}
\end{eqnarray}
Changing the variables one finds
\begin{eqnarray}
\label{first}
T=\frac{1}{L_x}\sum_n\int_{E_n(q)<k_0}\frac{dq}{2\pi}\sum_le^{-i(q-k_0sin\theta)al}\times\sum_me^{i(q-k_0sin\theta)am}\times\\ \nonumber
\times \int_0^adxe^{-iqx}u_{nq}^*(x) \int_0^adxe^{iqx}u_{nq}(x),
\end{eqnarray}
where wave functions in different layers of PC can be found from Eq.(\ref{sreq}) (see Appendix).
By substituting $\sum_ne^{-inqa}=2\pi\delta(qa)$ into Eq.(\ref{first}), for transmission coefficient we will have
\begin{equation}
T=\frac{1}{a}\sum_{E_n<k_0}\left|\int_0^au_n(x)e^{ik_0\sin\theta x}dx\right|^2
\label{final}
\end{equation}
where $u_n(x)\equiv u_{nq=k_0sin\theta}(x)$ . The sum in Eq.(\ref{final}) include states from different zones with both negative and positive energies with the same quasimomentum $k_0\sin\theta$. The number of terms in the sum depends on parameters $a,k_0,b$ and $\varepsilon$.

\section{Straightening of light}
Using formula from Supplement we numerically calculate the transmission coefficient of central diffracted wave for the following parameters $a=0.6\mu m, b=0.06\mu m,k_0=12\mu m^{-1}, \varepsilon=4$. The results are shown in  Fig.2 . We compare with the vacuum case $b\equiv 0$. In this case the sum Eq.(\ref{final}) contains only one term with $E_0=k_0\sin^2\theta/2$. Clear straightening effect is seen from Fig.2.
\begin{figure}
\vspace{-2cm}
 \begin{center}
\includegraphics[width=16.0cm]{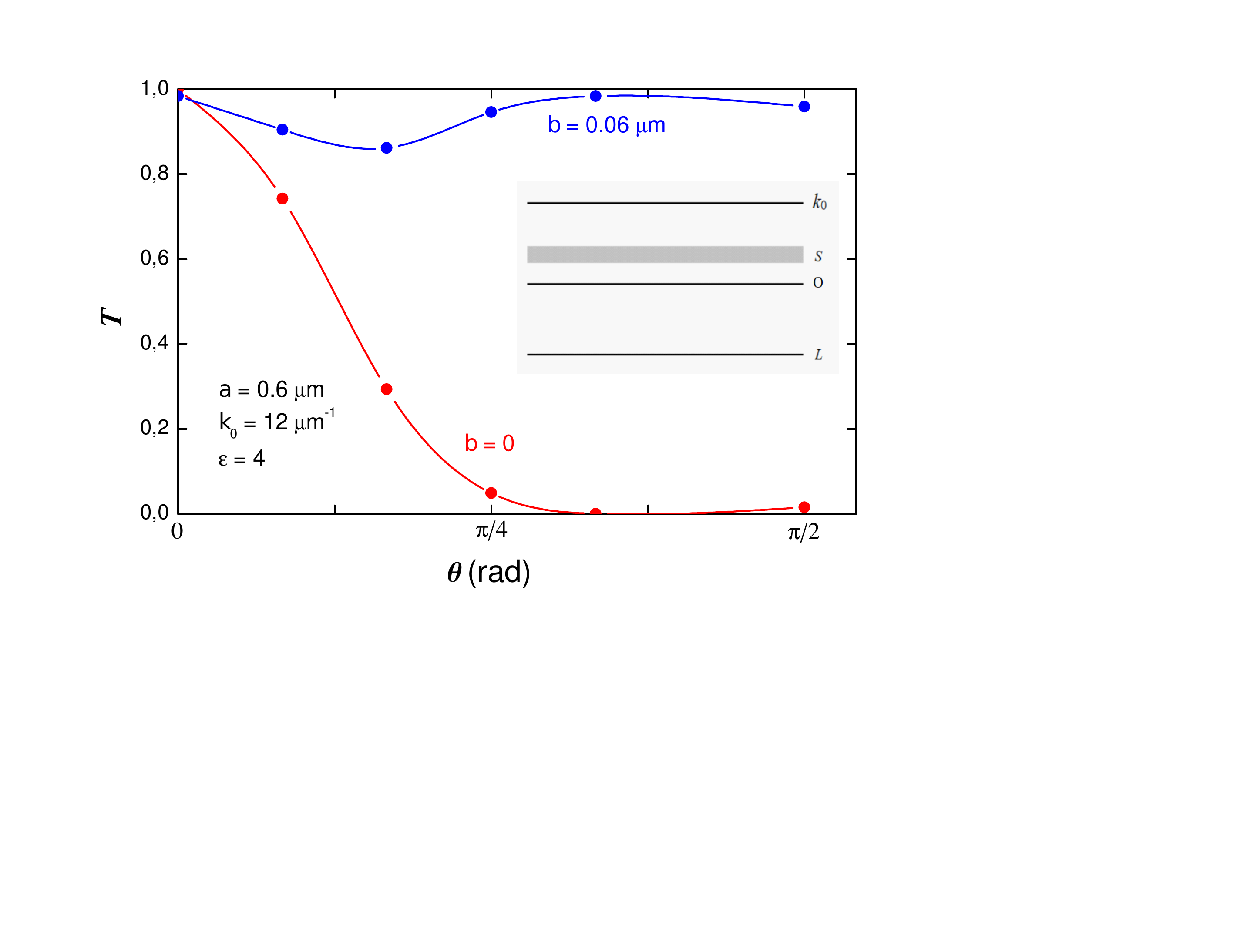}
\vspace{-4cm}
\caption{Transmission coefficient dependence on incident angle.Blue line is the theoretical plot with PC. Red line is the background theoretical plot without PC. In the inset the energy band scheme of transverse motion is shown.}
\label{fig.2}
\end{center}
\end{figure}

Indeed, independent of incident angle, light leaves the system mainly on the $0z$ direction because $T$ is close to unity.
It is clear, as mentioned above, that different energies give contribution to $T$. However, as indicate our numerical calculations, the main contribution to $T$ comes from relatively small positive $E_n>0$ and  relatively large negative $E_n<0$ states. 

The state, that close to the bottom of the conduction band, characterized by the small group velocity in transverse direction caused by flat dispersion curve and can lead to supercollimation effect, discussed in Refs. \cite{prather09}. As for the states with negative energies, they
form a very narrow band gap with almost localized states and with limited transport properties in transverse direction, see inset in Fig.2.

From theoretical point of view,
it is clear that under certain conditions  when the incident photon wavenumber is trapped in the gap of energy spectrum of
transversal motion, in DPCs light can propagate without spreading out.
These states also lead to the suppression of mobility in transverse direction and force photons move only parallel to plates. Note, that depending on the incident angle, the first or second type of states contribute differently to straightening effect, see Appendix. However, the resultant $T$, taking into account the contribution of all states (negative and positive) becomes almost incident angle independent and close to unity value.

\section{Conclusion and Discussion}

We have considered the transport of light through a $1d$ dilute photonic crystal model at oblique incidence. Our theoretical
study, based on Maxwell’s equations with a spatially dependent inhomogeneous
dielectric permittivity. For certain parameters' values of DPC the emerging light propagates in the normal direction despite the oblique incidence. This straightening effect is
intimately connected with limited transport properties of photons in transverse direction.

When considering the straightening effect we assume continuity of scalar wave field at the interface plane between two media (xy in Fig.1). Coming back to the em wave case this means that our consideration is correct for s-polarized waves, electric field vector of which is directed on 0y in geometry of Fig.1.
 As mentioned above the straightening effect in visible range could be utilized in solar cell elements to make their absorption efficiency higher. It could seem that the polarization dependence of the effect will decrease the application efficiency because the natural light consists of both s- and p-polarizations. However it follows from Fresnel formulae that the p-polarized wave is essentially reflected only for very large incident angles $\theta>70^0$. In contrary a s-polarized wave is reflected even for moderate incident angles. Therefore using DPC to straighten s-polarized light has sense.

As was mentioned above, in the discussed model of dilute photonic crystal, we have ignored the back-scattered waves and reflection.
In order to justify formally this approach, we
estimate below the value of reflection amplitude $r$, using the standard definition
 \begin{equation}\label{eff}
   r=\frac{\sqrt{\varepsilon_e}-1}{\sqrt{\varepsilon_e}+1}
 \end{equation}
where the effective dielectric constant $\varepsilon_e$ defined by $\varepsilon_e=(1-b/a)+\varepsilon b/a$.
Taking  $\varepsilon=4$ and dielectric  fraction $b/a\sim 0.1$, one has $\varepsilon_e\approx 1.3$ and $r\sim 0.06$,
i.e. neglecting back-scattered waves and reflection is a reasonable approximation while calculating $T$. As it follows from the above estimates a one-dimensional DPC on the one hand can straighten the oblique incident light in the visible range. On the other hand DPC provides a negligible reflection and
diffraction less light propagation, forecasting the potential of DPCs in increasing the absorption efficiency in solar cells. Note that a 1d photonic structure like Fig.1 could be made by litography on the surface of solar absorbing element at several wavelengths depth.


\section*{Acknowledgement}

 Zh.G. is grateful to Science Committee of Armenia for financial support (project 18T-1C082),A.Hakobian for help and A.Hakhoumian for useful discussions. E.C. and V.G. thank partial financial support by the Murcia Regional Agency of Science and Technology (project 19907/GERM/15). V.G. acknowledges the kind hospitality extended to him at Murcia University during his sabbatical leave.

\section*{Author contributions statement} Zh.G. and V.G. develope the theory, E.C. carry out the numerical calculations. All the authors review the manuscript.

\section*{Additional information}{\bf Competing Interests:} The authors declare that they have no competing interests.


\section{Appendix}
Bloch theorem states that the eigenstate $\phi_{\vec q}$ in a periodical potential can be represented in the form
\begin{equation}
\phi_{n\vec q}(\vec\rho)=e^{i\vec q\vec \rho}u_{n\vec q}(\vec \rho),
\label{eigenfu}
\end{equation}
where $u_{n\vec q}(\vec \rho)$ is a periodical function satisfying the equation
\begin{equation}
\left[-\frac{1}{2k_0}(i\vec q+\vec \nabla)^2+V(\vec \rho)\right]u_{n\vec q}(\vec\rho)=E_n(\vec q)u_{n\vec q}(\vec\rho).
\label{sreq}
\end{equation}
Below we present the solution of Scr$\ddot{o}$dinger equation Eq.(\ref{sreq}) in a unit cell and using it calculate the transmission coefficient. First consider positive energies $E>0$
\begin{eqnarray}
u_{q1}(x)=(A\cos\beta x+B\sin\beta x)e^{-iqx},\quad 0<x<a-b \nonumber \\
u_{q2}(x)=(C\cos\alpha x+D\sin \alpha x)e^{-iqx},\quad a-b<x<a
\label{wavefun}
\end{eqnarray}
with $\beta=\sqrt{2k_0E}$ and $\alpha=\sqrt{2k_0(V_d+E)}$. Remind that $V_d=k_0(\varepsilon-1)/2. $. Here we omit index n for simplicity.The constants $B, C, D$ can be expressed by $A$ using boundary conditions. $A$ itself can be found from the normalization condition $\int_0^a|u(x)|^2dx=1$.
Using continuity conditions and periodicity of wave functions one finds dispersion equation in the form
\begin{equation}
\cos (k_0a\sin\theta )=\cos\alpha b\cos\beta(a-b)-\frac{\alpha^2+\beta^2}{2\alpha\beta}\sin\alpha b\sin\beta(a-b)
\label{disp}
\end{equation}
and following relations between coefficients $C=A$, $D=\beta B/\alpha$ and
\begin{equation}
B=A\frac{\cos\alpha b-e^{ik_0\sin\theta}\cos\beta(a-b)}{e^{ik_0\sin\theta}\sin\beta(a-b)+\frac{\beta}{\alpha}\sin\alpha b}
\label{BA}
\end{equation}
Substituting Eq.(\ref{wavefun}) into Eq.(\ref{final}), using relations between coefficients and taking elementary integrals for a particular contribution of $E$ into transmission coefficient of central diffracted wave one has
\begin{equation}
T=T_1\times T_2
\label{teq}
\end{equation}
where
\begin{equation}
T_1=\frac{1}{ a}\bigg|\frac{\sin\beta(a-b)}{\beta}+\frac{\sin\alpha b}{\alpha}+\frac{\cos\alpha b-e^{ik_{0}asin\theta}\cos\beta(a-b)}{e^{ik_{0}asin\theta}\sin\beta(a-b)+\frac{\beta}{\alpha}\sin\alpha b}
\left(\frac{2\sin^2\frac{\beta(a-b)}{2}}{\beta}-\frac{2\beta\sin^2\frac{\alpha b}{2}}{\alpha^2}\right)\bigg|^2
\label{FA}
\end{equation}
and
\begin{eqnarray}
 \nonumber 
  T_2 =\bigg[\frac{a}{2}+\frac{\sin2\beta(a-b)}{4\beta}+ \frac{\sin2\alpha b}{4\alpha}+\left|\frac{\cos\alpha b-e^{ik_0asin\theta}\cos\beta(a-b)}{e^{ik_0asin\theta}\sin\beta(a-b)+\frac{\beta}{\alpha}\sin\alpha b}\right|^2\times\bigg.\\\nonumber
 \bigg.\times \left(\frac{a-b}{2}-\frac{\sin2\beta(a-b)}{4\beta}-\frac{\beta^2\sin2\alpha b}{4\alpha^3}+\frac{\beta^2b}{2\alpha^2}\right)\bigg.+\\\nonumber
   \bigg.+\frac{1}{2}\left( \frac{\cos\alpha b-e^{ik_0a\sin\theta}\cos\beta(a-b)}{e^{ik_0asin\theta}\sin\beta(a-b)+\frac{\beta}{\alpha}\sin\alpha b } + \frac{\cos\alpha b-e^{-ik_0asin\theta}\cos\beta(a-b)}{e^{-ik_0asin\theta}\sin\beta(a-b)+\frac{\beta}{\alpha}\sin\alpha b} \right)\times\\     \times \left(\frac{\sin^2\beta(a-b)}{\beta}+\frac{\beta\cos2\alpha b}{2\alpha^2}-\frac{\beta}{2\alpha^2}\right)\bigg]^{-1}
   \label{finalT}
\end{eqnarray}

Note that Eqs(\ref{teq}-\ref{finalT}) determine the contribution of a particular and positive solution of the dispersion equation Eq.(\ref{disp}) ($0<E<k_0$) into transmission coefficient. In order to find the total transmission coefficient one must, for given parameters $a,b,k_0$ and $\varepsilon$, find all positive solutions with $E < k_0 $ and sum up their contributions. Beside the mentioned positive solutions, one should take into account also the contribution of the negative solutions $E<0$. The dispersion equation and transmission coefficient for this case can be found from Eqs.(\ref{disp},\ref{FA},\ref{finalT}) by analytical continuation. The dispersion equation in this case acquires the form
\begin{equation}
\cos (k_0\sin\theta a)=\cos\alpha b\cosh\beta(a-b)-\frac{\alpha^2-\beta^2}{2\alpha\beta}\sin\alpha b\sinh\beta(a-b)
\label{dispneg}
\end{equation}
where $\beta=\sqrt{2k_0|E|}$,\quad $\alpha=\sqrt{2k_0(V_d-|E|)}$. Corresponding transmission coefficient has the form
\begin{equation}
T=T_1\times T_2
\label{teqn}
\end{equation}
where
\begin{equation}
T_1=\frac{1}{ a}\bigg|\frac{\sinh\beta(a-b)}{\beta}+\frac{\sin\alpha b}{\alpha}+\frac{\cos\alpha b-e^{ik_{0}asin\theta}\cosh\beta(a-b)}{e^{ik_{0}asin\theta}\sinh\beta(a-b)+\frac{\beta}{\alpha}\sin\alpha b}
\left(\frac{2\sinh^2\frac{\beta(a-b)}{2}}{\beta}-\frac{2\beta\sin^2\frac{\alpha b}{2}}{\alpha^2}\right)\bigg|^2
\label{FAneg}
\end{equation}

and
\begin{eqnarray}
\label{finalTneg}
\nonumber
 T_2 =\bigg[\frac{a}{2}+\frac{\sinh2\beta(a-b)}{4\beta}+ \frac{\sin2\alpha b}{4\alpha}-\left|\frac{\cos\alpha b-e^{ik_0asin\theta}\cosh\beta(a-b)}{e^{ik_0asin\theta}\sinh\beta(a-b)+\frac{\beta}{\alpha}\sin\alpha b}\right|^2\times\bigg.\\\nonumber
 \bigg.\times \left(\frac{a-b}{2}-\frac{\sinh2\beta(a-b)}{4\beta}+\frac{\beta^2\sin2\alpha b}{4\alpha^3}-\frac{\beta^2b}{2\alpha^2}\right)\bigg.+\\\nonumber
   \bigg.+\frac{1}{2}\left( \frac{\cos\alpha b-e^{ik_0a\sin\theta}\cosh\beta(a-b)}{e^{ik_0asin\theta}\sinh\beta(a-b)+\frac{\beta}{\alpha}\sin\alpha b } + \frac{\cos\alpha b-e^{-ik_0asin\theta}\cosh\beta(a-b)}{e^{-ik_0asin\theta}\sinh\beta(a-b)+\frac{\beta}{\alpha}\sin\alpha b} \right)\times\\     \times \left(\frac{\sinh^2\beta(a-b)}{\beta}+\frac{\beta\cos2\alpha b}{2\alpha^2}-\frac{\beta}{2\alpha^2}\right)\bigg]^{-1}
\end{eqnarray}
Now using the above mentioned expressions one can calculate transmission coefficient of central diffracted wave. Taking $a=0.6\mu m$,\quad $b=0.06\mu m$,\quad $\varepsilon=4$ \quad $k_0=12\mu m^{-1}$ and numerically calculating we get for $\theta=0$, $E_{s}=2.227$, $T_{s}=0.358$, $E_m=4.436$, $T_m=0$ $|E_b|=4.729$,$T_b=0.626$. Here $E_{s},E_m<12$ are the positive solution of dispersion equation Eq.(\ref{disp}), $T_{s,m}$ are the corresponding partial transmission coefficients calculated using Eq.(\ref{finalT}). Correspondingly $E_b,T_b$ are contributions from negative solution calculated using Eqs.(\ref{dispneg}) and (\ref{finalTneg}). The resulting transmission coefficient for incident angle $\theta=0$ is $T=T_{s}+T_{b}+T_m=0.984$. For any other incident angle the transmission coefficient can be calculated in analogous manner. We present results also for incident angle  $\theta=\pi/12$, $E_{s}=1.403$,$T_{s}=0.639$,\quad $E_m=6.259$,$T_m=0.008$ \quad $|E_b|=4.686$,$T_b=0.258$\quad $T=T_{s}+T_{b}+T_m=0.905$. For $\theta=\pi/4$, $E_s=1.748$, $T_s=0.496$,\quad $E_m=5.331$, $T_m=0.002$,\quad$|E_b|=4.708$, $T_b=0.448$\quad $T=T_s+T_m+T_b=0.946$. As it is obvious from these calculations the main contribution to the transmission coefficient of central diffracted wave give $s,b$ modes. The contribution of the $m$ mode with not small positive energy is negligible and not shown in band scheme Fig.3.The angle dependence of transmission coefficient is presented in Fig.2 of main text. Note that in the vacuum case $b=0$ there is no localized mode and positive energy as it follows from Eq.(\ref{disp}) equal $E_m=k_0\sin^2\theta/2$. We use this value when calculating transmission coefficient in vacuum case, see Fig.2.


\begin{thebibliography}{99}

\bibitem{joann08} Joannopoulos, J.D., Steven G.Johnson,S.G., Winn, J.N. and Meade,R.D., Photonic Crystals: Molding the Flow of Light (Princeton University Press, 2008).
\bibitem{john} Fink Y, Winn J N, Fan S, Chen C, Michen J, Joannopoulos J D and Thomas E L, A dielectric omnidirectional reflector,  Science {\bf 282}, 1679-1682, (1998).
\bibitem{oscar} Barco,O.Del,  Conejero, Jarque E, Gasparian, V. and Bueno,J.M. Omnidirectional high-reflectivity mirror in the 4-20 mum spectral range, J. Opt. {\bf 19}, 065102, (2017).
\bibitem{OMN2004} Onoda,M., Murakami,S. and Nagaosa,N., Hall Effect of Light,Phys.Rev.Lett.,{\bf 93},083901,(2004).
\bibitem{haldane08} Haldane,F.D.M. and Raghu,S., Possible Realization of Directional Optical Waveguides in Photonic Crystals with Broken Time-Reversal Symmetry, Phys.Rev.Lett. {\bf 100},013904,(2008).
\bibitem{Kos99} Kosaka,H., Kawashima,T.,  Tomita,A., Notomi,M.,Tamamura,T., Sato,T. and Kawakami,S.,Self-collimating phenomena in photonic crystals , Appl. Phys. Lett. {\bf 74}, 1212-1214(1999).
\bibitem{Kos00}Kosaka,H., Kawashima,T., Tomita,A., Sato,T. and Kawakami,S.,Photonic-crystal spot-size converter,  Appl. Phys. Lett. {\bf 76}, 268-270 (2000).
\bibitem{wu03}Wu,L.J., Mazilu,M. and Krauss,T.F. Beam steering in planar-photonic crystals: From superprism to supercollimator, J. Lightwave Technol. {\bf 21}, 561-566 (2003).
 \bibitem{prather04}Prather, D.W. and et al,Dispersion-based optical routing in photonic crystals, Opt.Lett. {\bf 29}, 50-52 (2004).
 \bibitem{pustai04}Pustai,D.M. ,Shi,S.Y., Chen,C.H. ,Sharkawy,A. and Prather,D.W., Analysis of splitters for self-collimated beams in planar photonic crystals, Opt. Express {\bf 12}, 1823-1831 (2004).
 \bibitem{shi04} Shi,S.Y., Sharkawy,A.,  Chen,C.H., Pustai,M.D.  and Prather,D.W., Dispersion-based beam splitter in photonic crystals, Opt. Lett. {\bf 29}, 617-619 (2004).
 \bibitem{rakich06} Rakich,P.T. and et al, Achieving centimetre scale super collimation in a large area 2D photonic crystal, Nature Materials, {\bf 5}, 93-96 (2006).
  \bibitem{miller06} Miller,A.B., Photonic crystals: Straightening out light, Nature Materials, {\bf 5},83-84,(2006).
\bibitem{prather09}Prather,D.W., Sharkawy,A., Shi,S., Schneider,G., Photonic Crystals, Theory, Applications and Fabrication,John Wiley and Sons,(2009).
\bibitem{aly12} Aly,A.H., Ismaeel,M., Abdel-Rahman,E.,Comparative Study of the One Dimensional Dielectric and Metallic Photonic Crystals, Optics and Photonics Journal,{\bf 2},105,(2012).
 \bibitem{zhao09} Zhao,J.,Li,X.,Zhong L., and Chen G.,Calculation of photonic band-gap of one dimensional photonic crystal, J.Phys.Conf.Ser.{\bf 183},012018,(2009).
  \bibitem{review}  Han,L., 1d Photonic Crystals: Principles and Applications in Silicon Photonics, DOI:10.5772/intechopen.7153; A.Vakhrushev,Theoretical Foundations and Application of Photonic Crystals, DOI:10.5772/intechopen.69145, (2018).
 \bibitem{GGC16}Gevorkian,Zh., Gasparian,V. and  Cuevas,E.,Bloch states in light transport through a perforated metal EPL, {\bf 113}, 64003 (2016).
\bibitem{gagc17}  Gevorkian,Zh., Hakhoumian,A., Gasparian,V.,Cuevas,E.,Capsize of polarization in dilute  photonic crystals Scientific Reports,{\bf 7},16593,(2017).
\bibitem{zhao2013}Zhao,J., Ruan,S.,Yan,P.,Zhang,H., Yu,Y.,Wei,H.,Luo,J., Cladding-filled graphene in a photonic
crystal fiber as a saturable absorber and its first application for ultrafast all-fiber laser,Optical Engineering, {\bf 52(10)},106105,(2013).
\bibitem{peig2015}Yan,P., Lin,Pg, Chen,H., Zhang,H., Liu,A., Yang,H., and Ruan,S.,Topological Insulator Solution Filled in Photonic Crystal Fiber for Passive Mode-Locked Fiber Laser,IEEE PHOTONICS TECHNOLOGY LETTERS, {\bf 27},264,(2015).
\bibitem{FF79} Feit,M.D.  and Fleck jr.,J.A.,Calculation of dispersion in graded-index multimode fibers by a propagating-beam method, Appl. Opt., {\bf 18}, 2843-2851 (1979)
\bibitem{Dyck85}D.Van Dyck , Advances in Electronics and Electron
Physics, Vol. 65 (Academic, New York) 1985, p. 295.
\bibitem{Lag89}De Raedt,H., Lagendijk,Ad,  and de Vries ,P., Transverse Localization of Light,Phys.
Rev. Lett., {\bf 62}, 47 (1989).
\bibitem{Krpen04}Kittel,C., Introduction to Solid State Physics (John Wiley and Sons) 2004.

\end{thebibliography}
\end{document}